\documentclass[aps,prl,reprint,groupedaddress,floatfix,10pt]{revtex4-1}	
\usepackage[latin9]{inputenc}
\usepackage{float}
\usepackage{amsmath}
\usepackage{amssymb}
\usepackage{graphicx}
\usepackage{esint}
\usepackage{color}

\makeatletter	
\graphicspath{ {/} }			
\usepackage{amssymb}

\begin{document}
\title{A Model of Energetic Ion Effects on Pressure Driven Tearing Modes in Tokamaks}
\author{M.R. Halfmoon}
\affiliation{Department of Physics and Engineering Physics, University of Tulsa, Tulsa, OK}
\author{D.P. Brennan}
\affiliation{Plasma Physics Laboratory, Princeton University, Princeton, NJ}

\date{\today}

\begin{abstract}
The effects that energetic trapped ions have on linear resistive magnetohydrodynamic (MHD) instabilities are studied in a reduced model that captures the essential physics driving or damping the modes through variations in the magnetic shear.  The drift-kinetic orbital interaction of a slowing down distribution of trapped energetic ions with a resistive MHD instability is integrated to a scalar contribution to the perturbed pressure, and entered into an asymptotic matching formalism for the resistive MHD dispersion relation.  Toroidal magnetic field line curvature is included to model trapping in the particle distribution, in an otherwise cylindrical model.  The focus is on a configuration that is driven unstable to the $m/n=2/1$ mode by increasing pressure, where $m$ is the poloidal mode number and $n$ the toroidal.  The particles and pressure can affect the mode both in the core region where there can be low and reversed shear and outside the resonant surface in significant positive shear. The results show that the energetic ions damp and stabilize the mode when orbiting in significant positive shear, increasing the marginal stability boundary.  However, the inner core region contribution with low and reversed shear can drive the mode unstable.  This effect of shear on the energetic ion pressure contribution is found to be consistent with the literature. These results explain the observation that the $2/1$ mode was found to be damped and stabilized by energetic ions in $\delta f$ - MHD simulations of tokamak experiments with positive shear throughout, while the $2/1$ mode was found to be driven unstable in simulations of experiments with weakly reversed shear in the core.  This is also found to be consistent with related experimental observations of the stability of the $2/1$ mode changing significantly with core shear.
\end{abstract}

\maketitle

\section{Introduction}\label{INTRO}
The onset of disruptive instabilities has long stood as a puzzling barrier to the realization of magnetic confinement fusion energy in tokamaks, by preventing access to the required high confined energy regimes.  As the toroidal current and confined energy are increased, a stability boundary can be crossed, typically for the $m/n=2/1$ tearing instability, which grows on a timescale slow with respect to the fast Alfv\'{e}n wave, but fast with respect to the global resistive time.  These tearing instabilities reconnect the magnetic field at magnetic surfaces with which the perturbation is resonant \cite{FKR}. Because the instability is small at the onset, the stability boundary and initial growth of the mode can be described by linearization of the equations to obtain an exponential growth. As the instability progresses, it forms magnetic islands which grow at an algebraic rate \cite{Rutherford}, and are nonlinear in their evolution at the minimum amplitude at which they are first detected \cite{Brennan05}. In fact, if the instability is marginally linearly stable, it can be easily destabilized into a nonlinear growth through coupling with another unstable mode, such as a sawtooth mode \cite{Brennan03}. As the mode grows to a significant island size, neoclassical effects can drive further growth \cite{Callen,Chang,Wilson,Maraschek}, effectively making the linear instability a seed mechanism for the nonlinear mode that is observed in experiment. Neoclassical tearing modes can limit the achievable $\beta\equiv\frac{P}{B^2}$, the ratio of pressure to magnetic field energy, and often cause disruptions \cite{Sauter}. Indeed, neoclassical tearing modes have recently been found to be the most likely cause of disruptions in JET \cite{deVries}, and are a primary cause of concern for upcoming burning plasma experiments on ITER.

Modern toroidal magnetic confinement devices achieve high $\beta$ by injecting beams of essentially monoenergetic ions that collide with the background plasma and slow down to produce a steady-state "slowing down" distribution of energetic ions in addition to the thermal distribution.  Burning plasma experiments such as ITER will also have a significant contribution from slowing down fusion alphas.  This type of energetic ion population has been shown to interact with a variety of magnetohydrodynamic (MHD) instabilities: resistive wall modes \cite{HuBetti}, sawtooth oscillations \cite{HuBettiManny}, toroidal Alfv\'{e}n eigenmodes \cite{Heidbrink}, and more recently resistive MHD modes \cite{Takahashi09a,Brennan12}.  
Understanding the physics driving the initial onset of the observed instabilities that cause disruptions can lead to their avoidance, and is of utmost importance in the development of a stable operational scenario for ITER.  Though ideal MHD stability with high energy ions has been thoroughly studied, the physics of energetic ion effects on resistive MHD instabilities has not been studied in great depth.  

Simulations of tokamaks have shown that energetic particles can drive the development of resistive instabilities as shown in Ref.~\cite{Brennan12}, but such particle contributions have also been shown to dampen the mode in simulations as in Ref.~\cite{Takahashi09a}. The discrepancy between these results presents a puzzle, and the analysis in this paper presents a possible solution. 

In Ref.~\cite{Takahashi09a} it was shown that there is a stabilizing effect from energetic ions for the slow growing $m/n=2/1$ tearing mode for equilibrium configurations with monotonic safety factor $q$ and strictly positive shear. In Ref.~\cite{Brennan12} it was found that the addition of energetic ions in weakly sheared or slightly reversed shear equilibrium configurations can drive $n=1$ instabilities. For low $q_{min}\gtrsim 1$, a non-resonant $m/n=1/1$ instability was driven by energetic ions with fishbone characteristics, where similar results have been found by other groups \cite{Feng}.  However, also in Ref.~\cite{Brennan12}, for $q_{min}\gtrsim 1.3$ a $2/1$ instability was driven by energetic ions. These results for the $2/1$ mode, between Refs.~\cite{Takahashi09a} and \cite{Brennan12}, would seem to conflict, given an intuition that the energetic ion contributions to the $2/1$ mode stability would dominantly originate from the vicinity surrounding the $q=2$ surface, which does not differ greatly in characteristics between the two cases.  However, this paper addresses how the differences in the equilibrium shear configurations in the core alter the contribution of energetic ions to the stability of the $2/1$ mode. As shown below, this effect can be significant, and can change their contribution from stabilizing to destabilizing to the mode.

Experimental results may also show similar characteristics.  In Ref.~\cite{Takahashi09a} it was conjectured that energetic particle effects may play a role in results from the JET experiment, where discharges have been found to be more stable to the $2/1$ mode than similar experiments on on DIII-D and JT-60U \cite{ButteryIAEA08}. Although the details of the shear profiles are not certain, these discharges were sawoothing, suggesting monotonic shear in the core between sawteeth.  Indeed, the most significant difference in the dimensionless parameters of these otherwise similar discharges is in the fraction of energetic ion pressure, where JET exceeds 30\% while DIII-D and JT-60U are typically below 20\%.  On the other hand, hybrid discharges in DIIID, on which the work of Ref.~\cite{Brennan12} was inspired, have flat or weakly reversed $q$ profiles, and the energetic ions appear to destabilize the $2/1$ mode in these discharges. 

If the instability grows to the neoclassical tearing mode stage, a number of interactions between high-energy particles and the mode can occur, as addressed in Refs.~\cite{Hegna,Mirnov,LaHaye06,Cai16}.  And, some of these interactions may bear some relation to the effects on the linear mode being discussed in this paper. In Ref.~\cite{Hegna} the effect of a passing distribution of energetic ions were found to be stabilizing to the island, if the density profile of energetic ions were controlled to increase outside the island location. Ref.~\cite{Mirnov} shows how FLR effects from energetic ions can interact with and stabilize nonlinear tearing modes in MST.  The work of Ref.~\cite{LaHaye06} predicts that locking to the wall can be mitigated by electron cyclotron current drive.  Ref.~\cite{Cai16} discusses how uncompensated cross field current can suppress neoclassical tearing modes in ITER for configurations with weak magnetic shear.  Effects such as these are relevant after the mode has grown to significant island size, whereas the focus of this paper is on the stability of the linear modes that can seed the neoclassical tearing mode.

In this paper, we have developed and applied a finite $\beta$ reduced MHD model which captures the physics of the trapped energetic ion effect on the linear growth of slow-growing tearing modes, as well as the ideal MHD stability limits.  In Ref.~\cite{HuBetti}, Hu and Betti showed that trapped energetic particles were stabilizing to resistive wall modes. Their formalism introduced kinetic effects to the eigenmode through a $\delta W_{K}$ calculation inserted into the Resistive Wall Mode (RWM) dispersion relation, and did not consider the effect of resonant surfaces and the tearing mode.  We extend this formalism into the model for the tearing mode, including asymptotic matching to the resonant surfaces, where resistivity and viscosity can be taken into account.  

We consider equilibria which are stable for zero pressure and driven unstable to the $2/1$ mode as pressure is increased. A perfectly conducting wall surrounds the plasma in this model, as it is beyond the scope of this work to consider the effects of a resistive wall and rotation, which will appear in future work.  Instead, the focus here is on the effect of shear on the stability of the mode. Our work has shown that the differences in the local magnetic shear can fundamentally alter the effects that particles have on tearing mode stability.

In Sec.~\ref{EQUIL} we describe the equilibrium configuration in detail, including how the equilibria are varied in the study.  In Sec.~\ref{IONS} we discuss the model used for high-energy ions: the physical significance of the "slowing down" distribution function used, the precession frequency of energetic ions, and the perturbed pressure contribution that arises from the non-Maxwellian particle population. In Sec.~\ref{TEARING}, we present the tearing mode stability analysis, and indicate how the energetic particle contribution enters into the stability of the mode through a perturbed contribution to the total pressure. In Sec.~\ref{RESULTS}, we show how the inclusion of energetic ions affects the marginal stability boundaries for a series of equilibrium configurations.  In Sec.~\ref{SUM}, we discuss the results, where we state in summary that with monotonic and significant shear the energetic ions are damping and stabilizing to the mode, whereas given a pressure gradient over weak shear in the core, the energetic ions can drive the $2/1$ mode unstable.

\section{The equilibrium configurations}\label{EQUIL}
We consider a simplified reduced MHD equilibrium configuration with step-function profiles in pressure and current density, similar to Refs.~\cite{Finn95,HuBetti,Brennan14}.  In Ref.~\cite{Brennan14} the current channel radius is set to $r_c=0.5$ and the pressure radius is set to $r_p=0.8$.  This pressure step is located just outside the rational surface location, influencing the mode to interact with the wall and have stability properties emulating the $2/1$ dominated and external-kink-like mode observed in experiment.  This configuration is used as a starting point for the analysis in this paper.  As mentioned in Sec.~\ref{INTRO}, the wall is considered to be perfectly conducting, with the focus on the interaction effects between tearing modes and energetic particles. Future work will include the addition of a resistive wall and coupling between the resistive wall mode and tearing modes similar to the work in Ref.~\cite{Brennan14} with the addition of kinetic ion effects.
\begin{figure}[t]
\centering
\includegraphics[width=3in]{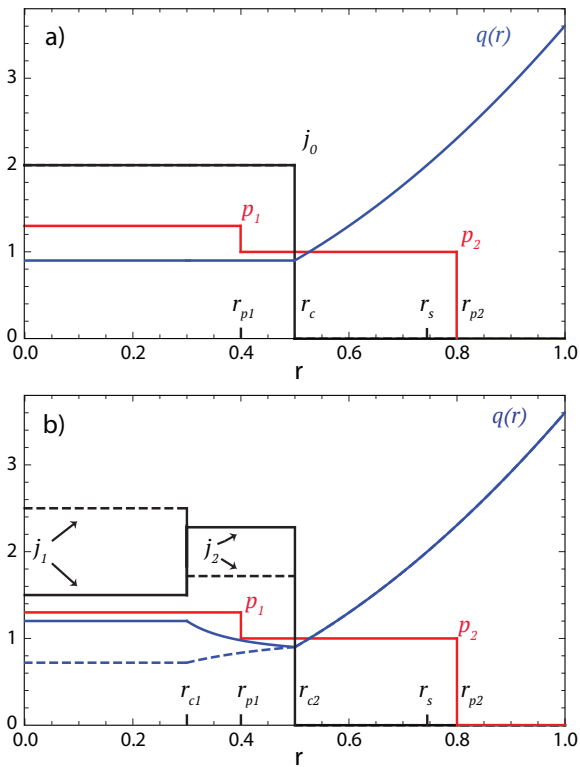}
\caption{The equilibrium profiles for the configuration with a single step in current a) and the case of two current steps b).  Both cases have two steps in pressure.  For the cases with a single current step, the shear at the internal pressure step is fixed at zero.  Two current steps allows for variable shear in the region of the internal pressure step by changing the amplitude of the core current, $j_1$, while the value of $j_2$ is set to keep the total current fixed. The solid (dashed) lines are for a configuration with negative (positive) shear.}
\label{fig:profs1}
\end{figure}

To examine the effect of the pressure contribution to the mode stability both in the low shear core and outer region in combination, two pressure steps are included as shown in Fig.~\ref{fig:profs1} a).  The first pressure step is placed at $r_{p1}=0.4<r_c<r_s$, where $r_s$ is the rational surface location. The second step is placed at $r_{p2}=0.8>r_s$ as in Ref.~\cite{Brennan14}.  The fractional difference $\delta_P\equiv(p_1-p_2)/p_1$ is held fixed at $\delta_P=0.3$ for most of the analysis unless otherwise noted, and is then varied to study the effect of pressure peaking on the drive to the mode.

Furthermore, to examine the effect of the shear and reversal in the $q$ profile on the stability, two steps are included in the current density function at $r_{c1}=0.3<r_{p1}$ and $r_{c2}$ with $r_{p1}<r_{c2}<r_{s}$, as shown in Fig.~\ref{fig:profs1} b). This places the first pressure step between the two current steps, to allow for control of the shear at the internal pressure step, with all three steps inside of the rational surface location. At radii below the first current step, $q\equiv r B_0/R B_{\theta}$ is constant. At radii outside of the outer current step $q$ increases monotonically $q\sim r^2$.  Between the two the shear $s\equiv (r/q)\partial q/\partial r$ depends on the two values $j_1$ and $j_2$, and can be reversed. Throughout this paper, the total current is held fixed as the $q$ profile is varied, making $j_2$ a function of $j_1$ and keeping the location of the rational surface fixed.  Also, the $q$ profile variation is limited such that $q(r<r_s)<q(r_s)$ for reversed shear, meaning that the only resonant location is at $r_s$, and we do not consider the double tearing mode. 

The pressure steps, $p_1$ and $p_2$, are fixed at $r = 0.4$ and $0.8$ respectively. The current steps, $j_1$ and $j_2$, are held at $r=0.3$ and $0.5$ respectively. The total current in each configuration is held constant, $I_{tot}=2\pi$, fixing the rational surface location at $r_s=0.745$. These positions are held constant to remove the effects of proximity of the pressure and current steps to the rational surface which can greatly affect stability.  The influence of the radial location of the rational surface, evident in the work of Furth Rutherford and Selberg\cite{FRS}, though physically present here, can be dominated by this proximity effect when using step function profiles.

All length scales are normalized to the wall radius such that $r_{w}=1$.  Time scales are normalized to $r_{w}/v_{A}$ where the Alfv\'{e}n speed $v_{A}$ is based on the poloidal field gradient on axis evaluated at the wall $B^\prime_{\theta0}(0)r_w$, which leads to normalization $B^\prime_{\theta0}(0)=r_w=1$.  The normalized equations for the equilibrium can then be written
\begin{align}
p_0(r)&=(p_1-p_2)\Theta(r_{p1}-r)+p_2\Theta(r_{p2}-r)\\
j_{z0}(r)&=(j_1-j_2)\Theta(r_{c1}-r)+j_2\Theta(r_{c2}-r)\\
B_{\theta0}(r)&=\frac{j_1}{2}r \, \qquad \qquad \qquad \qquad \ \ r<r_{c1} \nonumber \\
&=\frac{j_1r_{c1}^2+j_2(r^2-r_{c1}^2)}{2r}  \quad \ \ r_{c1}<r<r_{c2} &
\label{eqn:btheta}\\
&=\frac{j_1r_{c1}^2+j_2(r_{c2}^2-r_{c1}^2)}{2r} \ \ \quad r>r_{c2} & \nonumber \\
B_z&=B_0 \\
q(r)&=\frac{B_0}{R}  \qquad \ \ \ \qquad \qquad \qquad \ r<r_{c1} \nonumber &\\
&=\frac{B_0}{R_0}\frac{2r^2}{j_1r_{c1}^2+j_2(r^2-r_{c1}^2)}\ \ r_{c1}<r<r_{c2} \\
&=\frac{B_0}{R_0}\frac{2r^2}{j_1r_{c1}^2+j_2(r_{c2}^2-r_{c1}^2)}\ \ r>r_{c2}. \nonumber
\end{align}

Here, the major radius $R$ is assumed to satisfy $\epsilon\equiv r_w/R<<1$ and in reduced MHD, $B_{\theta0}\sim\epsilon B_0$ and $B_0$ is a constant. For a large aspect ratio with radially varying axial field $B_z(r)$, we can write $B_z B^\prime_z\sim B^2_{\theta}/r_w\sim \epsilon^2B^2_0$.  This means that at the steps in pressure and current density $r_{p1}$, $r_{p2}$, $r_{c1}$ and $r_{c2}$, it is consistent to have $B_0$ constant and neglect the explicit $J\times B=\nabla P$ equilibrium force balance, and yet maintain equilibrium force balance in the stability analysis.  

\section{The Model for Energetic Ions}\label{IONS}
The equilibrium model described above is based on a cylindrical reduced MHD configuration applicable to large aspect ratio tokamaks.  We now take into account the dynamics of high-energy ions to model their effect on mode stability.  The most relevant interaction is the resonance between the precessional drift frequency of trapped particles and the frequency of the mode. The precessional drift frequency of trapped particles is expressed as
\begin{equation}
\omega_D^{i,e} = -\frac{d\Phi}{d\Psi} + \frac{qv_{th}^2}{\Omega_cRr}H(u)\label{Eq:Frequency},
\end{equation}
where $\omega_E\equiv-\frac{d\Phi}{d\Psi}$ is the $E\times B$ drift with $\Phi$ the electrostatic potential and $\Psi$ the poloidal magnetic flux, and the second term on the right hand side represents the bounce averaged torodial drift of trapped banana orbiting particles. Here, $q$ is the safety factor, $v_{th}=\sqrt{2k T/m}$ is the thermal velocity, and as shown in Ref.~\cite{Roscoe} $H(u)=(2s+1)\frac{E(u)}{K(u)}+2s(u-1)-\frac{1}{2}$ is a function of complete elliptic integrals $E$ and $K$, the global magnetic shear ($s=-\frac{1}{q}r\frac{dq}{dr}$), and pitch angle of a given magnetic field ($u=Sin(\theta_B/2)^2$). For a standard trapped particle, $\omega_D\tau_A\sim 10^{-3}$. Due to the low frequency of the tearing mode $\omega\sim \eta <<\omega_B, \omega_E$, consistent with Ref.~\cite{HuBetti}, we assume the trapped particle resonant interaction occurs when $\omega_D - \omega \approx \omega_D \approx 0$. 

Given the temperatures and densities of the beam driven magnetic confinement experiments being addressed, we assume the ions are collisionless while the electrons are collisional on the bounce time.  Specifically, given that $\nu_e=2.9\times 10^{-12}n_e \lambda T_e^{-\frac{3}{2}}\approx 7.85 \times 10^8 T_e^{-\frac{3}{2}}s^{-1}$, $\nu_i=4.8\times 10^{-14}n_i \lambda \frac{m_i}{m_p}T_i^{-\frac{3}{2}}\approx 1.30 \times 10^7 T_i^{-\frac{3}{2}}s^{-1}$ for ITER-like values of $n_i=n_e\approx 10^{20}m^{-3}$, $\omega_D^{i,e}\approx 10^{-3}s^{-1}$ we find that $\omega_D > \nu_{eff}$ for $T_i > 5keV$ and $T_e > 35 keV$ where $\nu_{eff}=\frac{\nu^{i,e}}{\epsilon}$, $\epsilon$ is the aspect ratio. This leads to only ion pressure terms being retained for our study. 

Assuming that all beam ions enter isotropically at the birth energy with $v_{beam}$, we use the well known "slowing-down" equilibrium distribution function of the form
\begin{equation}\label{eq:3}
f_0(v)= \frac{P_0}{E_c^{3/2}+(\frac{1}{2}m_i v^2)^{3/2}}; v < v_{beam}.
\end{equation}
Here $P_0$ is a normalization constant, $v_{beam}$ is the injection beam velocity, and $E_{c}$ is the critical energy of the beam.   Using a "slowing-down" distribution is not only physically justified, but also gives this study a point of consistency with Refs.~\cite{Takahashi09a} and \cite{Brennan12} among many others. Our model normalizes $f_0(v)$ such that $P_0$ is set using $4\pi P_0\int_0^{v_{birth}}v^4f_0(v)dv/B_0^2=\beta_{i}\equiv\beta_0 \beta_{f}$ for a given $v_{birth}$ and $\beta_{f}$, where $\beta_i$ is the energetic ion $\beta$, and $\beta_f$ is its fraction relative to the equilibrium pressure.

Retaining only first-order perturbations, each quantity has a Fourier decomposition in the angular variables: $Q(r,\theta,\phi,t)=Q_0(r)e^{im\theta-in\phi}+\delta Q(r,\theta,\phi,t)$. The time-dependent component of the ion distribution function is a small-scale deviation from the steady-state function, calculated by 

\begin{equation}\label{eq:4}
\begin{split}
\frac{\partial\delta f}{\partial t} + (\mathbf{v_\parallel}+\mathbf{v_d})\cdot\nabla\delta f-\mu \nabla_\parallel B \frac{\partial\delta f}{\partial v_\parallel} \\ =-\mathbf{v_E}\cdot\nabla f_0 - (\frac{q}{m}E_\parallel +v_\parallel(\hat{b}\cdot\nabla\hat{b})\cdot\mathbf{v_E})\frac{\partial f_0}{\partial v_\parallel},
\end{split}
\end{equation}

\noindent from Ref.~\cite{Takahashi09a}, which reduces to:
\begin{equation}
(\omega+\omega_D^{i,e})\delta f = (\omega_N+(\hat{v}^2-\frac{3}{2})\omega_T+\omega_E)\frac{e \Phi f_0}{T_0}\label{Eq:deltaf},
\end{equation} 
assuming the characteristic frequencies describe the time-evolution of the distribution function. This form for $\delta f$ enters the ion pressure calculation assuming $\omega \approx 0$ as discussed above.

It is assumed the particles orbit in the equilibrium fields in this linear model.  To study their influence on the MHD stability, we include the mode structure into the calculation of the perturbed particle pressure.  The details of the solution of the MHD stability will be discussed in the next section.  For the present calculation of the perturbed particle pressure itself, it is sufficient to know the modes have spatial structure of $e^{im\theta + ikz}$ with $k=-n/R$ and $n=1$.

Magnetic curvature plays an important role in the dynamics of particle motion in tokamaks. In a cylindrical reduction, the field line curvature is captured by mathematical approximation. We start with the standard equation for magnetic curvature,
\begin{eqnarray}
\vec{\kappa}=\vec{b}\cdot\nabla\vec{b},
\end{eqnarray}
where $\vec{b}=\frac{\vec{B}}{|B|}$. Within the calculation of the energetic particle pressure, the radial component of the curvature primarily affects the field line displacement. Assuming a high aspect ratio, the radial component of the curvature can be expanded as an average and first order variational term. Assume the toroidal and poloidal curvature are modeled as:
\begin{align}
\vec{\kappa}_t=-\frac{\vec{R_0}}{R_0+rCos(\theta)}\\
\vec{\kappa}_p=-\frac{B_{\theta}^2}{rB_0}\hat{r}=\frac{r}{R_0q^2}\hat{r}.
\end{align}
We are primarily interested in the component of the curvature normal to the flux surface, $\vec{\kappa}\cdot\hat{r}$. For the toroidal component,

\begin{equation}
\vec{\kappa}\cdot\hat{r}=-\frac{1}{R_0}\frac{Cos(\theta)}{1+\frac{r}{R_0}Cos(\theta)}\approx -\frac{1}{R_0}(Cos(\theta)-\frac{1}{R_0}Cos^2(\theta))
\end{equation}

Averaging over the radial projection of both components, $\langle\vec{\kappa}\cdot\hat{r}\rangle$, and adding the first order poloidally varying term, $\vec{\tilde{\kappa}}$, we obtain the approximation $\vec{\kappa}(\theta)=\langle\vec{\kappa}\cdot\hat{r}\rangle+\vec{\tilde{\kappa}}$ 
\begin{equation}
\kappa(\theta)=\frac{r}{R_0^2}(\frac{1}{2}-\frac{1}{q^2})-\frac{Cos(\theta)}{R_0},
\label{eq:kappa}
\end{equation}
or more compactly, $\kappa(\theta)=\kappa_0+\kappa_1 Cos(\theta)$.

The pressure contribution from energetic ions is calculated by taking the solution to the drift-kinetic equation Eq.~\ref{Eq:deltaf}, and taking a moment over this distribution,
\begin{equation}
\tilde{p}^m_i=\int \int v^4 \delta f(v) dv^3 dx^3.
\end{equation}
The energetic ions follow along magnetic field lines with bounce points determined by conservation of energy. This determines the ion precession frequency in Eq.~\ref{Eq:Frequency} as detailed in Ref.~\cite{Roscoe}. The pitch angle variable, $u=1-(R/r)(1-\mu B/ \epsilon)$, is a normalized description of this ion path. This variable sets the spatial limits in the pressure integral.
The poloidal harmonics, $Y_l^i$, are governed by the field line displacement and perturbed potential and describe the interaction that particles have with each mode. 
\begin{equation}
Y_l^i=\int_{-\pi}^\pi\frac{d\theta}{2\pi}e^{-il\theta}(\hat{v}^2\mathbf{\tilde{\xi}_\perp}\cdot\mathbf{\kappa}+\frac{eZ_i}{T_i}(\tilde{Z}+\mathbf{\tilde{\xi}_\perp}\cdot\nabla\Phi)).
\label{eq:Y}
\end{equation}
The drift orbit interactions with the magnetic field are described by the $\sigma$ functions in Eq.~\ref{eq:sigma}, which are a combination of complete elliptic integrals that describe the resonance between modes and particles:
\begin{equation}
\sigma_m =\int_0^{\pi/2}d\chi\frac{Cos[2(m-q(r))Sin^{-1}(\sqrt{u}Sin(\chi))]}{K(u)\sqrt{1-uSin^2(\chi)}}.
\label{eq:sigma}
\end{equation}

The combination of these terms results in the particle pressure integral of Ref.~\cite{HuBetti}, where a Maxwellian background distribution was assumed. For a Maxwellian in the presence of a step function equilibrium, the kinetic pressure contribution is written as 
\begin{align}
\tilde{p}^m_i=\frac{2^{5/2}\epsilon^{1/2}}{5\pi^{3/2}}\int_0^{\infty}d\hat{v}^5e^{-\hat{v}^2}\int_0^1duK(u)\Pi_i\sigma_m\sum_{l=-\infty}^{+\infty}\sigma_lY_l^i\nonumber\label{eq:ptilde}\\
\end{align} 
In this paper, we assume a slowing down distribution function.  Using this assumption with the step function equilibrium configuration in the calculations for the energetic particle pressure component, we substitute the curvature from Eq.~\ref{eq:kappa} into Eq.~\ref{eq:Y} and reduce the series of poloidal integrals into Dirac delta distributions,
\begin{equation}
\begin{split}
\tilde{p}_i^m=\int_0^{v_{max}}d\hat{v}^5 f_0(\hat{v}^2)\int_0^1 du K(u)\sigma_m\Pi_i \times \\ (\sum_{l=-\infty}^\infty\sigma_l (\frac{\hat{v}^2}{B_0rR_0}\kappa_1\frac{m}{2}\tilde{\psi}_m(r_p)(\delta_{m+1,l}+\delta_{m-1,l})
\\ +\frac{\hat{v}^2}{B_0rR_0}\kappa_0m\tilde{\psi}_m(r_p)\delta_{m,l}+\frac{Z_ieu_{z0}}{T_i}\tilde{\psi}_m(r_p)\delta_{m,l}\\+\frac{Z_ie}{T_i}\frac{mu_{z0}}{h_m(r_p)B_0}\tilde{\psi}_m(r_p)\delta_{m,l})),
\label{eq:pressure}
\end{split}
\end{equation}
where the total perturbed particle pressure is $\tilde{p}^K_i=\sum_m \tilde{p}_i^m e^{im\theta}$ with $\hat{v}=\frac{v}{v_{th}}$, $h_m=1-m/q_m(r)$, $r_p$ is the radial location of the equilibrium pressure step, and $Z_i$ is the ion charge number. $\Pi_i$ describes the ratio of the ion frequencies to the drift frequency that gauges the equilibrium pressure contribution to the perturbed pressure at any location   
\begin{equation}
\Pi_i=p_0 \frac{\omega_E + \omega_{*N}^i +(\hat{v}-3/2)\omega_{*T}^i }{\omega_D},\label{eq:pi}
\end{equation}
where $\omega_E$ is the $\mathbf{E} \times \mathbf{B}$ drift frequency, and $\omega_{*N}$ is the ion drift frequency. 

Equations \ref{eq:ptilde} and \ref{eq:pressure} are linear in the equilibrium pressure step amplitude and perturbation amplitude $\tilde{\psi}_m(r_p)$ at the pressure step.  The integral in Eq.~\ref{eq:pressure} is numerically evaluated to arrive at a function $\Lambda_i$ that depends only on $E_c$ and $v_{birth}$, in the perturbed energetic particle contribution:
\begin{equation}
\tilde{p}^m_i = C \Lambda_i(E_c,v_{birth}) \beta_f \psi_m(r_p). 
\end{equation}
The result is a scalar modification to the total perturbed pressure: $\tilde{p}=\tilde{p}_F+\tilde{p}^K$, where our calculations only consider the $2/1$ mode, and $\tilde{p}^K=\tilde{p}_i^m e^{im\theta}$. This total perturbed pressure is substituted into the outer region stability equation.
Next, we will apply the pressure contribution modeled here to the stability analysis of the 2/1 tearing mode.

\section{The Stability Model}\label{TEARING}
We compute the stability of the step function equilibria defined in Sec.~\ref{EQUIL} by asymptotic matching of the ideal outer region perturbation to a resonant layer model at the rational surface $r_s$.  The outer region model is derived from finite $\beta$ reduced ideal MHD without inertia.  Magnetic perturbations are described in terms of the perturbed flux $\tilde\psi\equiv\tilde A_z$, with $A_z$ the parallel component of the vector potential.  The perpendicular velocity perturbation is written in terms of a stream function $\phi$ as $v_\perp=\nabla\phi\times \hat{e}_z$, with $\hat{e}_z$ the unit vector. The outer region equations become
\begin{eqnarray}
\centering
0&=&iF(r)\nabla^2_\perp\tilde{\psi}-\frac{im}{r}j_0'(r)\tilde{\psi}+\frac{2imB_{\theta0}^2(r)}{B_0^2r^2}\tilde{p} \label{eq:outer}\\
\gamma\tilde{\psi}&=&iF(r)\tilde{\phi} \label{eq:flux_growth}\\
\gamma\tilde{p}_F&=&-\frac{im}{r}p^\prime_0\tilde{\phi}\label{eq:fluidpressure},
\end{eqnarray}
where $F(r)\equiv mB_{\theta0}(r)+kB_0$.  The step function characteristic of the equilibrium fields leads to Eq.~\ref{eq:outer} evaluating as $\nabla_\perp^2\tilde{\psi}=0$ for every radial position other than the equilibrium profile jumps in current and pressure:
\begin{equation}
\begin{split}
\label{eq:PsiTilde}
\nabla^2_\perp \tilde{\psi}=-[A_1\delta(r-r_{c1})+A_2\delta(r-r_{c2}) \\ +B_1\delta(r-r_{p1})+B_2\delta(r-r_{p2})]\tilde{\psi}
\end{split}
\end{equation}
where 
\begin{eqnarray}
A_i&=&\frac{2m}{r_{ci}F(r_{ci})}\\
B_i(\beta_f)&=&\frac{mB_{\theta0}(r_{pi})^2\beta}{r_{pi}^2F(r_{pi})}(\frac{m}{r_{pi}F(r_{pi})}-C_i\Lambda_i(\beta_f))
\label{Beq}
\end{eqnarray}
and $i\in [1,2]$. 

The perturbed magnetic flux solution $\tilde \psi$ has the standard cylindrical $r^{\pm m}$ form, and can be written as a linear combination of basis functions (in radial order) 
\begin{equation}
\tilde{\psi}=a_{c1}\phi_{c1}+a_{p1}\phi_{p1}+a_{c2}\phi_{c2}+a_s\phi_s+a_{p2}\phi_{p2}.
\end{equation}
where each basis function is peaked at an equilibrium step function location or the rational surface ($s$), and goes to zero at the next neighboring equilibrium step or surface (See Ref.~\cite{Brennan14}:Appendix for a detailed explanation).  
Each basis function has a radial structure dependent only on geometry on either side ($\pm$) of the step function or surface it represents.  For example, at the first pressure step, the two sides of the basis function are defined as
\begin{eqnarray}
\phi^{-}_{p1}(r)=\frac{(\frac{r}{r_{c1}})^m-(\frac{r_{c1}}{r})^m}{(\frac{r_{p1}}{r_{c1}})^m-(\frac{r_{c1}}{r_{p1}})^m}\\
\phi^{+}_{p1}(r)=\frac{(\frac{r}{r_{c2}})^m-(\frac{r_{c2}}{r})^m}{(\frac{r_{p1}}{r_{c2}})^m-(\frac{r_{c2}}{r_{p1}})^m}.
\end{eqnarray}
The linear combination of these basis functions, substituted into the eigenfunction solution, given the normalization that the flux at the rational surface is fixed at one ($\tilde{\psi}(r_s)=a_{s}=1$). The jumps in the derivatives of the basis functions become driving terms in the stability analysis, and are given by

\begin{eqnarray}
\left[\phi'_{c1}\right](r_{c1})=-\frac{2m}{r_{c1}}\frac{(\frac{r_{c1}}{r_{p1}})^m}{(\frac{r_{p1}}{r_{c1}})^m-(\frac{r_{c1}}{r_{p1}})^m}\label{stabini}
\\
\left[\phi'_{p1}\right](r_{p1})=-\frac{2m}{r_{p1}}\frac{(\frac{r_{p1}}{r_{c1}})^m}{(\frac{r_{c1}}{r_{p1}})^m-(\frac{r_{p1}}{r_{c1}})^m}
\end{eqnarray}

\begin{eqnarray}
\gamma\tau_t\tilde{\psi}(r_s)=\left[\tilde{\psi}'\right]_{r_s}\label{surfjump}
\\ \left[\tilde{\psi}'\right]_{r_{ci}}=-A_i\tilde{\psi}(r_{ci})
\\ \left[\tilde{\psi}'\right]_{r_{pi}}=-B_i(\beta_f)\tilde{\psi}(r_{pi}).\label{stabfin}
\end{eqnarray}
with $i\in [1,2]$ and $\gamma$ the growth rate of the mode.  In Eq.~\ref{surfjump} we have implemented a viscoresistive inner layer model at the rational surface for simplicity, with layer timescale $\tau_t$, though in principle more detailed physics models at the surface could be considered. Defining the usual stability index from Eq.~\ref{surfjump} as the jump in the logarithmic derivative across the surface $\Delta^\prime\equiv\left[\tilde{\psi}'\right]_{r_s}/\tilde{\psi}(r_s)$, we can solve for the stability of the mode in terms of $\Delta^\prime$. For this study, we focus on the marginal stability boundaries, where $\Delta^\prime=0$, and their variation under the influence of energetic ion pressure contributions. The result is a system of five equations to be solved for the five unknowns $a_{c1}, a_{p1}, a_{c2}, a_{p2},$ and $\Delta^\prime$ with $a_s=1$. For now, we ignore the effect of a finite mode frequency in the resistive layer and operate under the constant-$\psi$ approximation in our tearing calculations.  Additional effects, such as the interaction of precession frequencies and response frequencies at the rational surface, are a subject of future work as discussed below in Sec.~\ref{SUM}.

To illustrate the method we initially solve for $\Delta^\prime$ with a single current step at $r_{c2}$, ie.~$j_1=j_2$, and a single pressure step at $r_{p2}$, ie.~$p_1=p_2$.  In this case the stability is equivalent,  without energetic particles, to the resistive plasma - ideal wall result of Ref.~\cite{Brennan14}.
Solving Eqs.~\ref{stabini}-\ref{stabfin} for $\Delta'$ following the method described in the Ref.~\cite{Brennan14} Appendix, we find a simple form for $\Delta^\prime$ as a function of the equilibrium quantities, and note that it is now dependent on the fraction of high-energy trapped ions through the function $B$ in Eq.~\ref{Beq}. For the single step equilibrium configuration
\begin{equation}
\Delta ^\prime=\delta_t + \frac{l_{tc}l_{ct}}{\delta_c+A}+\frac{l_{pt} l_{tp}}{\delta_p+B_p(\beta_f)}.
\label{eq:Delta_p_single}\end{equation}
Where the inductance coefficients, $l_{ij}$, are defined in Sec.~\ref{APP}, Eq.'s A1-A8, as the derivatives of the basis flux functions at their boundaries and $\delta_i$ are the discontinuities at the peaks of each corresponding basis function, given by Eq.'s A9-A13. While the calculations in the Appendix are specifically for the case of the double-step configuration, one can easily apply them to the case of single current and pressure steps.

When we include two current and two pressure steps, the derivation leads to a $\Delta'$ of the form:
\begin{equation}
\begin{split}
\label{eq:Delta_p}
\Delta'=
\delta_t-\frac{l_{tc} l_{ct}}{(\delta_{c2}+A_2-\frac{l_{cp2}l_{pc2}}{(\delta_{p1}+B_{1p}-\frac{l_{pc1}l_{cp1}}{\delta_{c1}+A_1})})}
\\- \frac{l_{tp}l_{pt}}{\delta_{p2}+B_{2p}}
\end{split}
\end{equation}
For details of this derivation, see the Appendix. We note that the energetic ions enter the stability equation in two places, $B_{1p}$ and $B_{2p}$. In both Eq.~\ref{eq:Delta_p_single} and \ref{eq:Delta_p} the energetic ion effect on the stability at the outer pressure step is the same ($B_{2p}$), and depending on the sign of $\Lambda_i$ in Eq.~\ref{Beq}, can either increase or decrease the denominators in which they appear.  The energetic ion contribution at the inner pressure step has a more complicated structure, but on careful analysis can be seen to affect the mode in much the same way. 

The magnetic shear enters into the particle pressure contribution in the $\Pi_i$ terms of Eq.~\ref{eq:pi} for each individual jump. As discussed below in some detail, the amplitude of this term is governed by the particle precession frequencies for the population that exists at that radial location.  We now examine the effect the energetic ions have on the stability as the internal shear is varied.

\begin{figure}[ht]
\centering
\includegraphics[width=3in]{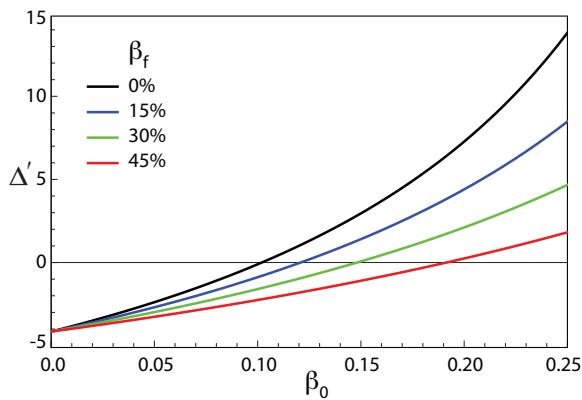}
\caption{$\Delta'$ values for several values of $\beta_f$. $\beta$ stability limit is increased for increasing $\beta_f$. This shows that energetic particle interactions are stabilizing to the 2/1 tearing mode for the case of single pressure and current steps.}
\label{fig:delta-prime}
\end{figure}

\begin{figure}
\centering
\includegraphics[width=3in]{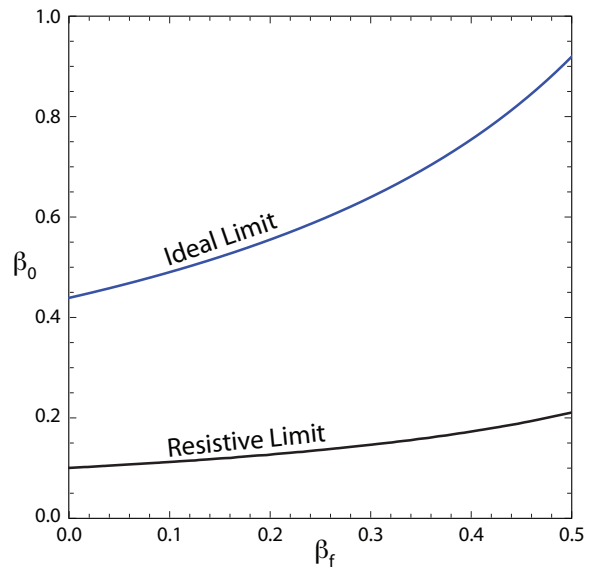}
\caption{Marginal stability ($\Delta'=0$) point and Ideal Stability ($\Delta'=\infty$) limit as a function of $\beta_0$ and $\beta_f$. Note that the highest stable $\beta_0$ increases with increasing $\beta_f$. This indicates that the particle pressure contribution has a stabilizing effect on the 2/1 tearing mode.} 
\label{fig:damped}
\end{figure}

\section{Energetic Ion Effects on the Tearing Mode}\label{RESULTS}
We first present results for the case of single pressure step and single current.  In Fig.~\ref{fig:delta-prime} $\Delta^\prime$ as a function of $\beta_0$ is shown in the case of single steps for a set of energetic ion $\beta_f=(0,0.15,0.30,0.45)$. As the energetic ion density is increased, we see the energetic ions have a stabilizing effect on the $2/1$ resistive mode, as indicated by a reduction of $\Delta^\prime$ and an increase in the stability boundary in $\beta_0$.  This case is used as a verification benchmark to ensure the $\beta$ stability limits are consistent with previous results and between Eqs.~\ref{eq:Delta_p_single} and \ref{eq:Delta_p}, and note that the marginal stability point at $\beta_f=0$ corresponds to the resistive plasma - ideal wall limit in Fig.~2 of Ref.~\cite{Brennan14}.

Both the ideal and resistive stability boundaries are affected by the energetic ion population. The ideal stability boundary can be identified as the location in equilibrium parameter space where $\Delta^\prime\rightarrow\infty$ as the denominators in Eqs.~\ref{eq:Delta_p_single} and \ref{eq:Delta_p} go to zero. Using this condition to solve for the ideal stability boundary, Fig.~\ref{fig:damped} shows both the resistive and ideal stability boundaries as a function of $\beta_0$ and $\beta_f$ for the single step case.  The stabilizing influence of the energetic ions on both is clear.

We now consider cases with two steps in both pressure and current.  With two steps in equilibrium pressure, we use a volume averaged $\beta_{eff}\equiv\int\beta dV/V$ to characterize the stability, as opposed to $\beta_0$. This volumetric averaging gives the expression

\begin{equation}
\beta_{eff}=\frac{\beta_0(1-\delta_p+\frac{r_{p1}^2}{r_{p2}^2})}{1+\frac{r_{p1}^2}{r_{p2}^2}},
\end{equation}
upon integration, where $\delta_p$ is the percent drop in pressure at the first step.

\begin{figure}
\centering
\includegraphics[width=3in]{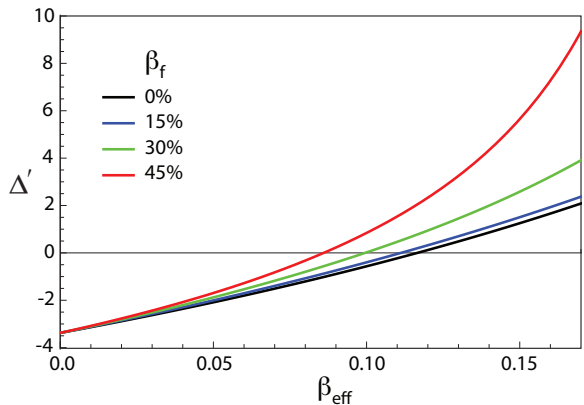}
\caption{$\beta$ stability limit decreases for increasing values of $\beta_f$ in the case with an internal pressure step in the negative shear region. This indicates that energetic particles are driving the 2/1 tearing mode unstable for this case. This equilibrium has a negative internal shear due to an internal current density jump. ($s=-0.277$)}
\label{fig:reverseshear}
\end{figure}

With the internal current $j_1=1.7$, which specifies a $j_2>2$ to keep the total current unchanged, and a fractional pressure difference of $\delta_P=0.3$, the lowest stable $\beta_{eff}$ with no particle effects present occurs at $\beta_{eff}=0.118$ as shown in Fig.~\ref{fig:reverseshear}.  In this case, as the particle fraction is increased, the mode becomes destabilized by the energetic ions.  This is evident by the increased $\Delta^\prime$ throughout and the reduction in the marginal stability boundary in $\beta_{eff}$. This destabilization is caused by the energetic ion contribution at the internal pressure step at $r_{p1}$, corresponding to the second term on the right hand side of Eq.~\ref{eq:Delta_p}.  The contributions of the energetic ions at $r_{p2}$, corresponding to the third term on the right hand side of Eq.~\ref{eq:Delta_p}, remain qualitatively unchanged from the results shown in Figs.~\ref{fig:delta-prime} and \ref{fig:damped}.

\begin{figure}[H]
\centering
\includegraphics[width=3in]{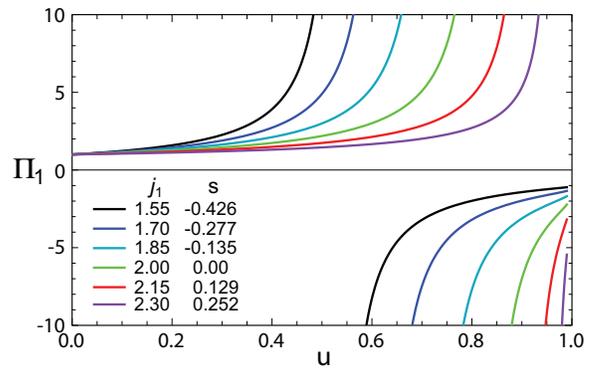}
\caption{As the shear increases, the resonant pitch angle increases. For the reversed shear cases, this resonance drives the pressure contribution to the tearing mode towards instability(Left). For a case with positive shear (Right), the energetic particle pressure contribution does not resonate and thus does not drive the mode.}\label{fig:Pi}
\end{figure}

This qualitative difference in the affect on the stability of the mode due to changes in the internal pressure and current structure can be understood in terms of the analytic model for the energetic ion pressure contribution to the stability through $B_{pi}$. The precession frequency and $\Pi_i$ are functions of pitch angle variable $u$ and are integrated over $u$ for the pressure contribution.  For typical values of magnetic shear, as the pitch angle is varied, the precession frequency goes from positive to negative, and gives a singular (but regular) contribution to $\Pi_i$ at the zero point as shown in Fig.~\ref{fig:Pi}.  The integration over $u$ gives a finite $\Pi_i$ despite the pole due to the symmetric opposing contributions on either side.  Fig.~\ref{fig:Pi} shows $\Pi(u)_i$ at the inner pressure step for a series of equilibria with values of $j_1$ varying from $1.55$-$2.3$ with steps of $0.15$ with fixed total current as in Fig.~\ref{fig:profs1} b). As the shear is varied, the location of the pole in $u$ changes, and the integral changes sign.  For energetic ion contributions at pressure steps in large positive shear, as is the case at $r_{p2}$, the precession frequency is mostly in the direction of the toroidal magnetic field leading to a positive integral and a stabilizing effect.  For large negative shear regions, which occurs at $r_{p1}$, the counter-field precession leads to a negative integral of $\Pi_i$, and a destabilizing influence on the mode. The resulting stability is a combination of the two influences, as described by Eq.~\ref{eq:Delta_p}.

\begin{figure}[H]
\centering
\includegraphics[width=3in]{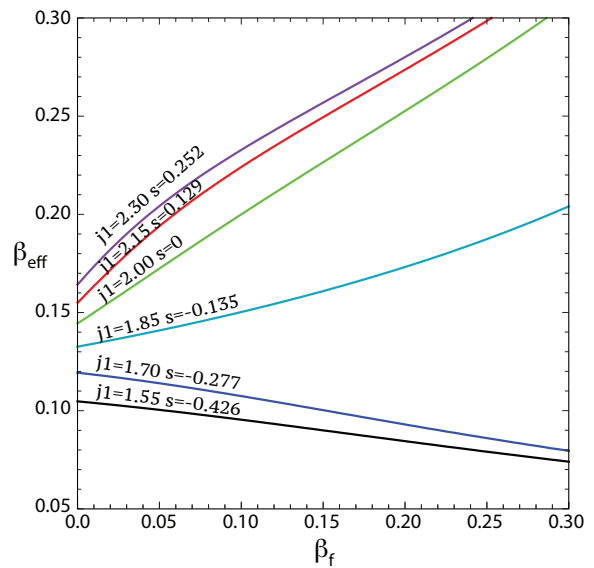}
\caption{The effect ions have on the marginally stable ($\Delta'=0$) $\beta_0$'s are dependent on the value of magnetic shear at the internal pressure step. The shear is dependent on the amplitude of the internal current jump. Values of $j_1$ vary from (top to bottom) 2.3-1.55 with steps of 0.15}
\label{fig:doublecontour}
\end{figure}

Figure \ref{fig:doublecontour} shows the shear dependence of the marginal stability point for the double current double pressure equilibrium configuration. Each line represents the $\Delta^\prime=0$ crossing for the given shear and internal current value, $\beta_{eff}$ values above this line are unstable and values below are stable. Observe that the $\beta_f=0$ crossing changes for
\noindent each value of the shear, this is due to the change in the internal equilibrium current value altering the stability criterion.

As Fig.~\ref{fig:doublecontour} shows, decreasing the shear alters the effect of the addition of particles from stabilizing to destabilizing. This decrease in the marginally stable $\beta_{eff}$ is also shown in Fig.~\ref{fig:reverseshear}. The destabilization caused by particles arises from the resonant contribution to the perturbed pressure that dominates for highly negative shear. The location of the resonant pitch angle depends directly on the shear as shown in Fig.~\ref{fig:Pi} due to the denominator of Eq.~\ref{eq:pi}.

\begin{figure}
\centering
\includegraphics[width=2.75in]{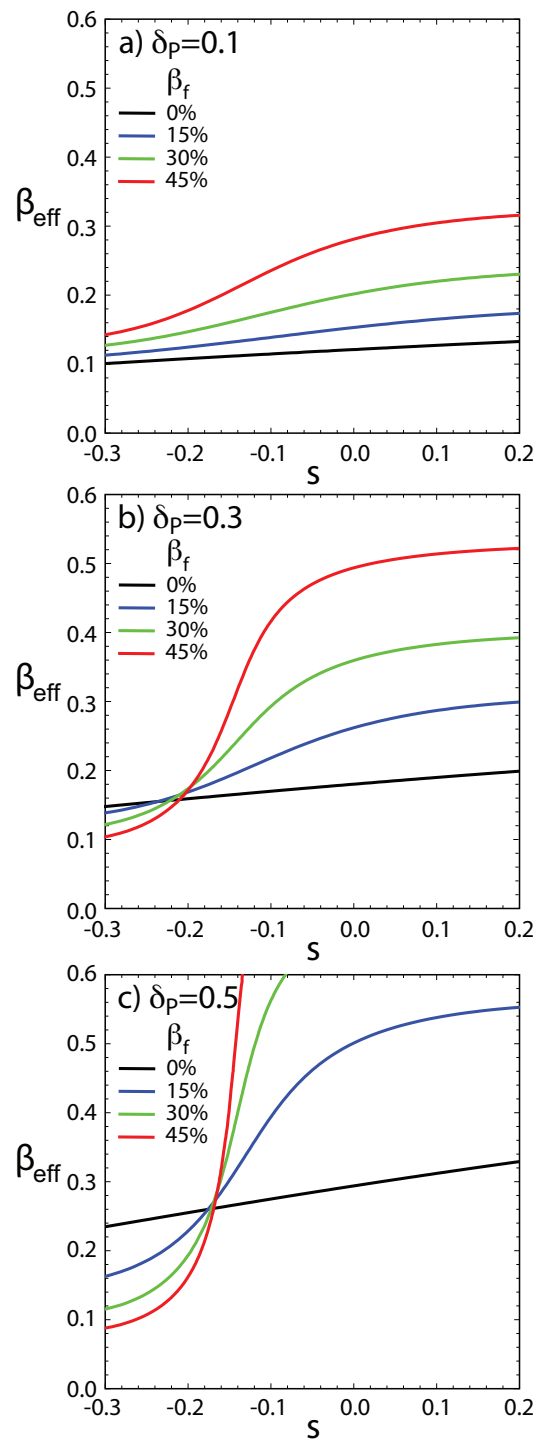}
\caption{The marginal stability boundaries ($\Delta'=0$) as a function of shear at the internal pressure jump $s$ and $\beta_{eff}$ for a series of energetic particle $\beta_f$ and internal pressure steps $\delta_P$ of $0.1$ a), $0.3$ b) and $0.5$ c). Above each line in $\beta_{eff}$ the mode is unstable.  For equilibria with significant reversals in $q$ ($s<-0.2$), and significant pressure contribution in the reversal region $\delta_P\gtrsim 0.25$, particles contribute a destabilizing effect to the 2/1 tearing mode. For equilibria with $s\gtrsim -0.2$, particles mitigate the growth of the 2/1 tearing mode.}
\label{fig:shear}
\end{figure}

In Fig.~\ref{fig:shear} is shown the marginal stability boundaries ($\Delta'=0$) as a function of shear at the internal pressure jump $s$ and $\beta_{eff}$ for a series of energetic particle $\beta_f$ and internal pressure steps $\delta_P$ of $0.1$ a), $0.3$ b) and $0.5$ c).  Above each line in $\beta_{eff}$ the mode is unstable, and below each line the mode is stable.  For each value of $\delta_P$, which effectively varies the pressure peaking factor, the current $j_1$ is varied over the range $1.7\lesssim j_1 \lesssim 2.3$ to give the the shear at the inner step ranging over $-0.3 < s < 0.2$. For a smaller internal pressure step $\delta_P=0.1$ the drive to the mode from the particles is dominated by the outer pressure step, and the stability boundaries increase with $\beta_f$ for all values of shear.  As the internal pressure step is increased to $\delta_P=0.3$, while at positive internal shear the mode is strongly stabilized, at negative values of internal shear ($s\lesssim -0.2$), the mode becomes destabilized by the internal pressure drive.  For a larger internal pressure step $\delta_P=0.5$, the mode becomes strongly stabilized at positive shear and destabilized at negative shear, showing a significant impact on the stability of the mode from the internal shear.  

It should be noted that in Fig.~\ref{fig:shear} the value of $\beta_{eff}$ at which the mode becomes unstable is affected by $\delta_P$ even in the absence of energetic particles.  So, some care should be taken when quantitatively comparing the $\beta_{eff}$ plots to the single pressure step equilibrium.

Note also that experimentally relevant values of $\delta_P$, taken by simply considering the pressure profile gradients in the equilibria shown in  Refs.~\cite{Brennan12} and \cite{Takahashi09a}, are probably closer to $0.5$ than $0.1$.  Though this study was conducted with a cylindrical model, the qualitative behavior observed can be used to interpret toroidal simulation results in accurate experimental configurations.  The $q$ profile in Ref.~\cite{Takahashi09a} is monotonically increasing with positive shear throughout.  In this case the energetic ions were found in the simulations to be robustly stabilizing to the $2/1$ mode.  In Ref.~\cite{Brennan12} the $q$ profile was weakly reversed.  And, despite similar equilibrium characteristics otherwise to the equilibria in Ref.~\cite{Takahashi09a}, the particles were found to be robustly destabilizing to the $2/1$ mode (note the region at $q_{min}>1.3$ in Ref.~\cite{Brennan12}).  From the work presented here, it is evident that the internal shear is responsible for this destabilization, in particular because of the destabilizing influence of the reversed precessing trapped particles within the distribution.

\section{Discussion and Conclusions}
\label{SUM}
In this paper we have shown analytically how the global magnetic shear governs the interactions between energetic ions and linear resistive tearing modes by determining whether the kinetic pressure contribution is stabilizing or destabilizing. The equilibrium configuration allows for an analysis of the effect of shear at different radial locations on the trapped particle pressure contribution to the stability of the resistive mode. In this case, the low or reversed shear near the core leads to a destabilizing influence in the dispersion relation, which can overcome the robustly stabilizing influence of the positive shear regions at outer radii. In summary with monotonic and significant shear the energetic ions are damping and stabilizing to the mode, whereas given a pressure gradient over weak shear in the core, the energetic ions can drive the $2/1$ mode unstable.  This effect originates from the resonance of the trapped particle precession frequency with the low-frequency tearing mode. This precession frequency is determined by the value of the shear for all allowable pitch angles, and critical values in the shear serve as turning points for the interaction effects from high-energy ions. This result is consistent with the trends in stability in simulations of reversed shear configurations as compared to configurations that are sheared to the core, and suggests this physics is driving the difference in the simulation results.

The results presented in this paper are focused on the energetic ion interaction with the ideal outer region, and how that can affect the drive to the stability of the resistive plasma mode.  In that respect, many important effects in the model for the equilibrium, plasma response and surrounding wall have been ignored to simplify the analysis and allow for this focus. We now discuss some of these physical effects that have been ignored, which could be important under certain conditions, and yet if introduced into this model would make the result  quantitatively, or even qualitatively, more accurate. 

In this study, all calculations for particle motion are done in the plasma frame, thus the mode rotation is set to zero. Adding in the effects of rotation and a resistive wall would allow us to directly compare with the analysis of the RWM from Ref.~\cite{HuBetti} and progress beyond those results, such as developing a model of coupling between tearing and resistive wall modes in the presence of energetic ions.  Analytically, in our configuration the stability equation is simply a scalar function with a matching condition, $\gamma\tau_A=\Delta^\prime$, whereas including a resistive wall and rotation would introduce a matrix equation with the off-diagonal terms which describe the coupling to the wall.  The relative frequencies of plasma rotation and ion precession, in the lab frame relative to the stationary wall, must all then be considered.  To be consistent in the energetic ion pressure calculation, the rotation also enters Eq.~\ref{eq:pressure}, where $u_{z0}$ is the relative equilibrium velocity of the core (considered to be zero for all calculations in this study).  This inclusion of equilibrium flow is left for future work.

In Ref.~\cite{Cai11}, the effect of circulating energetic ions in the form of a drift current perturbation leads to stabilization of tearing modes in the instance of co-circulation and destabilization for counter-circulating. The additional momentum provided by energetic ions leads to interactions with the outer region of tearing modes.  In our study this effect is analogous to the effect captured in Eq.~\ref{Eq:Frequency}, specifically in the $H(u)$ formula.  However, in our study, only trapped ion contributions are considered.  Although the trapped ion contribution is known to be dominant, both populations significantly contribute to the overall mode stability \cite{kim08}.  Ion driving and damping effects are governed by the precession frequency of particles relative to the mode frequency in both cases.

A number of effects can cause a finite frequency response in the resonant layer \cite{Finn16}, including two-fluid species, toroidal curvature and parallel dynamics, all of which can alter the stability and mode locking boundaries.  This can be particularly important when the equilibrium rotation is finite and different regimes of plasma response are traversed \cite{Cole06}.  However, the physics of the shear effects on the interaction with the energetic ions presented in this paper, and the resulting trends in stability, should remain intact.

While the kinetic effects of energetic electrons on the layer response \cite{DrakeLee} can be important in high temperature tokamaks, in particular  when the plasma enters the semi-collisional and collisionless regimes, the kinetic ion interaction presented in this paper is dominantly effective in the ideal outer region.  This stems from the mass ratio of the electrons to ions, and the spatial scales of their orbits at the relevant temperatures.  The kinetic layer response could thus be introduced as a resonant layer model, separate, but potentially consistent with the kinetic ion distribution.

The linear stability of the $2/1$ tearing mode can also be affected by coupling to other poloidal components in toroidal geometry, notably the $1/1$ and $3/1$ mode components \cite{Brennan06}.  Our model does not include this coupling in the MHD stability of $2/1$ tearing mode directly, but does take this into account in the calculation of the particle pressure contribution through additional modes in the $\sigma$ functions. The poloidal harmonics in Eq.~\ref{eq:Y} reduce to Dirac delta functions which model the effect that modes of differing $m$'s have on one another.  Future work will investigate the effect of poloidal mode component coupling on the stability analysis directly, rather than only that occurring in the energetic particle analysis, i.e. a matrix solution for the tearing and interchange eigenvalues. 

\section*{Acknowledgments}
We would like to thank John Finn, Andrew Cole and Dov Rhodes for the multitude of insightful comments over the course of this work, and their contribution to the development of our analytic model.
This work of was supported by the DOE Office of Science, Fusion Energy Sciences
under Grants No.~DE-SC0004125 and DE-SC0014005. 

\appendix
\label{APP}
\section{Solving $\Delta'$ by the Method of Auxiliary Functions}
The total perturbed magnetic flux is solved for by breaking the radial domain into sections separated by the discontinuities caused by changes in equilibrium quantities. Each sub-domain is described by a single auxiliary function. The homogeneous auxiliary function equation in cylindrical coordinates is 
\begin{equation*}
\frac{1}{r}\frac{\partial}{\partial r}r\frac{\partial\phi(r)}{\partial r}-\frac{m^2}{r^2}\phi(r)=0,
\end{equation*}
due to the poloidal dependence of perturbed quantities.

This leads to a general solution of the form:
\begin{equation*}
\phi_i(r)=Ar^m+Br^{-m}
\end{equation*}
These functions are normalized to one at the discontinuities in perturbed magnetic flux, and zero at the boundaries of their domain; the adjacent discontinuous radial locations.

The induction coefficients $l_ij$ take the form of radial derivatives of the auxiliary functions, $\phi_i$. The amplitudes of the discontinuities in the basis functions are represented by $\delta_i$

\begin{align}
\label{eq:Induction}
l_{cp1}=-\phi_{c1}'(r_{c1}+)\\
l_{pc1}=\phi_{p1}'(r_{c1}-)\\
l_{sc}=\phi_s'(r_{c2}-)\\
l_{cs}=-\phi_{c2}'(r_{s}+)\\
l_{sp}=-\phi_{s}'(r_{p2}+)\\
l_{ps}=\phi_{p2}'(r_{s}-)\\
l_{cp2}=\phi_{c2}'(r_{p1}-)\\
l_{pc2}=-\phi_{p1}'(r_{c2}+)\\
\delta_{c1}=[\phi'_{c1}]_{r_{c1}}\\
\delta_{p1}=[\phi'_{p1}]_{r_{p1}}\\
\delta_{s}=[\phi'_{s}]_{r_{s}}\\
\delta_{c2}=[\phi'_{c2}]_{r_{c2}}\\
\delta_{p2}=[\phi'_{p2}]_{r_{p2}}
\end{align}

Equation \ref{eq:PsiTilde} is solved at each of the discontinuous radial locations
\begin{eqnarray}
\label{eq:Jumps}
[\tilde{\psi'}]_{r_{c1}}=-A_1\tilde{\psi}(r_{c1})\\
{[\tilde{\psi'}]}_{r_{p1}}=-B_1\tilde{\psi}(r_{p1})\\
{[\tilde{\psi}]}_{r_{c2}}=-A_2\tilde{\psi}(r_{c2})\\
{[\tilde{\psi'}]}_{r_s}=\Delta'\tilde{\psi}(r_s)\\
{[\tilde{\psi'}]}_{r_{p2}}=-B_2\tilde{\psi}(r_{p2})
\end{eqnarray}

This leads to a linear combination of induction coefficients and equilibrium quantities in the solution for $\Delta'$.

\begin{eqnarray}
a_{c1}\delta_{c1}+a_{p1}l_{pc1}=-A_1a_{c1}\\
a_{p1}\delta_{p1}+a_{c1}l_{lcp1}+a_{c2}l_{cp2}=-B_1a_{p1}\\
a_{c2}\delta_{c2}+a_{p1}l_{pc2}+a_{s}l_{tc}=-A_2a_{c2}\\
a_{s}\delta_{s}+a_{c2}l_{cs}+a_{p2}l_{ps}=\Delta' a_{s}\\
a_{p2}\delta_{p2}+a_{s}l_{sp}=-B_2 a_{p2}
\end{eqnarray}

Algebraically solving this system for $\Delta'$ gives Eq.~\ref{eq:Delta_p}, given that $a_s$ is normalized to unity.

\end{document}